\documentclass[conference]{IEEEtran}
\IEEEoverridecommandlockouts

\usepackage{cite}
\usepackage{amsmath,amssymb,amsfonts}
\usepackage{algorithmic}
\usepackage{graphicx}
\usepackage{textcomp}
\usepackage[bookmarks=false]{hyperref}
\hypersetup{
    colorlinks = true,
    citecolor  = blue,
    linkcolor  = blue,
    urlcolor   = blue,
}
\usepackage{xcolor}
\usepackage{booktabs}
\usepackage{multirow}
\usepackage{pifont}
\usepackage{makecell}

\def\equationautorefname~#1\null{Eq.~(#1)\null}
\def\figureautorefname~#1\null{Fig.~#1\null}
\def\tableautorefname~#1\null{Tab.~#1\null}
\def\definitionautorefname~#1\null{Def.~#1\null}
\def\sectionautorefname~#1\null{Sec.~#1\null}
\def\subsectionautorefname~#1\null{Sec.~#1\null}
\def\subsubsectionautorefname~#1\null{Sec.~#1\null}

\def\BibTeX{{\rm B\kern-.05em{\sc i\kern-.025em b}\kern-.08em
T\kern-.1667em\lower.7ex\hbox{E}\kern-.125emX}}

\begin{document}

% =========================
\title{Accuracy-Configurable Floating-Point Multiplier Design for SRAM-Based Compute-in-Memory

\thanks{
This work was supported by the Fundamental Research Funds for the Central Universities (Grant Nos. 30925010605 and 30924012004), and the Jiangsu Provincial Major Science and Technology Project (Grant No. BG2025012).

$^{\dagger}$ These authors contributed equally.
$^*$ Corresponding authors.
}

}

\author{
\IEEEauthorblockN{
Yiqi Zhou$^{\dagger}$,
Junhao Lu$^{\dagger}$,
Jiale Yu,
Zhuo Xu,
Yang He,
Yue Yuan,
Shan Shen$^{*}$ and Daying Sun$^{*}$
}

\IEEEauthorblockA{
\textit{Nanjing University of Science and Technology, Nanjing, 210094, China} \\
Email: \{zhouyiqi, lujunhao0516, 923101760434, 125132011891,\\ 125132011874, yuanyue1801, shanshen, hasdysun\}@njust.edu.cn
}
}

\maketitle

% =========================
\begin{abstract}
Digital Compute-in-Memory (DCiM) reduces data movement and has become a promising solution for energy-efficient edge AI. However, most existing DCiM frameworks still primarily target integer or fixed-point arithmetic, and provide limited support for compiler-integrated and accuracy-configurable floating-point computation. Directly integrating conventional IEEE~754 floating-point units into dense SRAM-based DCiM arrays, however, incurs high area and power overhead. To address this challenge, this work presents an accuracy-configurable floating-point multiplier integrated into the OpenACM framework for SRAM-based DCiM. An exact IEEE~754-compliant multiplier is first implemented as a baseline, and a mantissa-segmentation-based approximate multiplier is then proposed to reduce hardware cost while preserving numerical fidelity. Post-layout results show up to 69\% logic area reduction and 72\% power savings over exact floating-point designs without delay overhead. Evaluations on image processing tasks and ResNet-18 inference further demonstrate negligible accuracy degradation. These results indicate that compiler-integrated approximate floating-point multiplication is a practical approach for enabling efficient and configurable floating-point support in SRAM-based DCiM systems. The Floating-Point Multiplier is available on \href{https://github.com/ShenShan123/OpenACM}{GitHub}.
\end{abstract}


\section{Introduction}
The rapid growth of data-intensive artificial intelligence workloads has intensified the memory wall bottleneck in conventional von Neumann architectures, where frequent data transfers between memory and processing units dominate energy consumption. Digital Compute-in-Memory (DCiM) mitigates this challenge by performing arithmetic operations directly inside memory arrays, thereby reducing data movement and improving energy efficiency. As a result, SRAM-based DCiM has emerged as a promising solution for edge neural network inference and energy-constrained intelligent systems.

However, most practical DCiM designs, including OpenACM~\cite{openacm}, primarily target integer or fixed-point arithmetic, and their support for floating-point computation remains limited. Although several recent CiM/DCiM prototypes have demonstrated the feasibility of floating-point or mixed integer/floating-point execution through carefully customized datapaths~\cite{isscc24_dualmode_gccim,isscc24_fp_reram,isscc25_fp_sramcim}, existing SRAM-based DCiM frameworks still lack general, compiler-integrated, and accuracy-configurable support for floating-point computation. In particular, floating-point operators are rarely exposed as first-class design components for systematic exploration, and configurable precision as well as approximate floating-point arithmetic are generally unsupported. These limitations significantly constrain the design space of SRAM-based DCiM and hinder efficient system-level exploration of accuracy--PPA trade-offs.

A straightforward way to introduce floating-point capability into DCiM is to integrate conventional IEEE~754 floating-point units around memory arrays. However, such an approach is difficult to scale in dense SRAM-based DCiM because of the substantial area and power overhead of standard floating-point hardware. Therefore, the practical challenge is not merely to enable floating-point computation in DCiM, but to support it in an efficient and configurable manner under tight power and area budgets.

Approximate computing provides a practical solution to this challenge by introducing \emph{controllable} arithmetic error to reduce circuit complexity. In floating-point computation, approximate floating-point multipliers (AFPMs) exploit the non-uniform significance of mantissa bits to simplify partial-product generation and accumulation, thereby reducing area and power with limited impact on computation quality~\cite{afpm_log_eafpu,afpm_static_seg,afpm_mmbs}. More importantly, such approximation naturally enables configurable accuracy--PPA trade-offs, which are highly desirable for DCiM design-space exploration. In SRAM-based DCiM, where a large number of multipliers are instantiated around memory arrays, even modest per-multiplier savings can accumulate into substantial system-level gains in energy efficiency and scalability.

Despite this potential, existing AFPM designs are usually evaluated as standalone arithmetic units, while most DCiM compilation frameworks remain centered on integer arithmetic and lack native support for compiler-integrated floating-point configuration and approximate floating-point operators. As a result, arithmetic-level advances in AFPM design cannot be readily translated into system-level DCiM optimization. To bridge this gap, this work introduces both exact and approximate floating-point multipliers into the OpenACM~\cite{openacm} flow, enabling systematic exploration of floating-point support and accuracy--PPA trade-offs within the compiler flow for SRAM-based DCiM.

The main contributions of this work are summarized as follows:
\begin{itemize}
    \item We implement an \textit{exact} IEEE~754-compliant floating-point multiplier as a correctness-preserving baseline following the standard multiplication pipeline.
    
    \item We design an accuracy-configurable \textit{approximate} floating-point multiplier based on mantissa segmentation and lightweight compensation, enabling fine-grained accuracy--PPA trade-offs.

    \item We integrate both exact and approximate floating-point multipliers into the OpenACM~\cite{openacm} DCiM compilation framework, exposing floating-point precision as an optimization knob within the compiler flow.
\end{itemize}
\section{Background and Motivation}
\label{sec:background}

Approximate floating-point multipliers (AFPMs) have been widely studied to reduce the hardware cost of floating-point computation while preserving acceptable numerical accuracy. Existing AFPMs can be broadly divided into logarithmic-based and segmentation-based methods.

Logarithmic AFPMs transform multiplication into addition using approximations such as $\log_2(1+x)\approx x$. Li \emph{et al.} proposed an accuracy-configurable AFPM that combines logarithmic approximation with multi-level error compensation and runtime reconfiguration, achieving substantial area reduction while maintaining low average relative error~\cite{afpm_log_eafpu}. Despite their high efficiency, logarithmic-based AFPMs often exhibit non-uniform error distributions and degraded accuracy for small operands, which may adversely affect numerical robustness.

Segmentation-based AFPMs explicitly exploit the non-uniform significance of mantissa bits. Di Meo \emph{et al.} proposed a static-segmentation AFPM that restructures mantissa multiplication into a multiply-and-accumulate form, eliminating costly shifters and achieving significant area and power savings with strong numerical robustness in image processing workloads~\cite{afpm_static_seg}. Li \emph{et al.} further introduced a mixed-precision segmentation scheme with runtime reconfigurable precision and frequency, showing that segmentation-based AFPMs can provide fine-grained accuracy--efficiency trade-offs while maintaining stable behavior in neural network applications~\cite{afpm_mmbs}. Compared with logarithmic approaches, segmentation-based AFPMs generally provide more predictable error characteristics and are therefore more suitable for energy-efficient floating-point DCiM design.

Although these works establish AFPMs as efficient arithmetic primitives, they are mostly evaluated in isolation and are not designed for compiler-driven DCiM integration. Existing DCiM compilation frameworks have also largely focused on integer arithmetic and inference-oriented kernels. General-purpose frameworks such as AutoDCIM~\cite{autodcim2023}, DAMIL-DCIM~\cite{damildcim2025}, OpenC$^2$~\cite{openc2_2025}, and OpenACM~\cite{openacm} do not provide native floating-point support. Although recent works such as ARCTIC~\cite{arctic2024}, SynDCIM~\cite{syndcim2024}, and SEGA-DCIM~\cite{segadcim2025} introduce floating-point capability, they are still restricted to fixed formats such as FP8/16/32 and rigid architectures. Even specialized high-precision solutions such as Hy-FPCIM~\cite{ma2025high} and MDCIM~\cite{liu2023mdcim} do not expose accuracy-configurable AFPMs as compiler-integrated design options, limiting their flexibility for hardware--software co-optimization.

\autoref{tab:dcim_compiler_fp} summarizes the floating-point and approximation support of representative DCiM compilation frameworks. Existing flows generally neither support arbitrary floating-point bitwidths nor expose accuracy-configurable AFPMs as compiler-integrated design choices. OpenACM~\cite{openacm} established an open-source, accuracy-aware compilation flow for SRAM-based approximate DCiM architectures with full physical design support, enabling systematic exploration of \emph{integer} approximate operators. However, its original operator library is restricted to integer arithmetic and does not support native floating-point multiplication.

Therefore, the key missing piece is compiler-integrated \emph{approximate} floating-point support. This work addresses that gap by incorporating exact floating-point multipliers as correctness baselines and accuracy-configurable AFPMs as energy-efficient alternatives into the OpenACM flow, enabling systematic co-optimization of numerical format, operator selection, and hardware cost.

\begin{table}[tbp]
\centering
\caption{Comparison of DCiM compilation frameworks with respect to floating-point capability and approximation support.}
\label{tab:dcim_compiler_fp}
\resizebox{\linewidth}{!}{%
\begin{tabular}{lccc}
\toprule
Framework & Precision Support & FP Bitwidth & Multiplier Architecture \\
\midrule
AutoDCIM~\cite{autodcim2023}        
& INT 
& \textcolor{red}{\ding{55}} 
& \textcolor{red}{\ding{55}}  \\

ARCTIC~\cite{arctic2024}          
& INT/FP  
& FP8/16/32/BF16 
& Shared-exponent  \\

SynDCIM~\cite{syndcim2024}         
& INT/FP 
& FP8/BF16 
& Bit-wise CSA  \\

SEGA-DCIM~\cite{segadcim2025}       
& INT/FP 
& FP8/16/32/BF16  
& Pre-aligned FP \\

DAMIL-DCIM~\cite{damildcim2025}      
& INT 
& \textcolor{red}{\ding{55}} 
& \textcolor{red}{\ding{55}}  \\

OpenC$^2$~\cite{openc2_2025}       
& INT 
& \textcolor{red}{\ding{55}} 
& \textcolor{red}{\ding{55}} \\

OpenACM~\cite{openacm} 
& INT 
& \textcolor{red}{\ding{55}} 
& \textcolor{red}{\ding{55}} \\

Hy-FPCIM~\cite{ma2025high} 
& FP 
& FP16/BF16 
& Booth Mantissa Macro \\

MDCIM~\cite{liu2023mdcim} 
& FP 
& FP64
& Modified FMA \\

\midrule
This work       
& \textcolor{green!60!black}{\checkmark} 
& \textcolor{green!60!black}{\checkmark} 
& FP8-32/AFP16-32 \\
\bottomrule
\end{tabular}%
}
\vspace{-12pt}
\end{table}

% =========================
\section{OpenACM with Floating-Point Support}
\label{sec:floating-point}
The discussion above highlights a critical gap between recent advances in
approximate floating-point multiplier design and their practical deployment in
DCiM systems. While AFPMs offer compelling accuracy--efficiency trade-offs at
the arithmetic level, their adoption in SRAM-based DCiM architectures requires
careful integration with array-level execution models and compilation flows.
To bridge this gap, we incorporate both exact and approximate floating-point
multipliers into the OpenACM DCiM framework, enabling floating-point operators
to be instantiated, configured, and evaluated within a compiler-driven design
flow.

\subsection{Exact Floating-Point Multiplication}
The exact floating-point multiplier described in this subsection follows the
IEEE~754 multiplication pipeline and serves as a correctness-preserving
baseline for evaluating the proposed approximate designs in DCiM settings. For any floating-point number compliant with the IEEE 754 standard, its binary representation consists of a sign bit $S$, an exponent field $E$, and a mantissa field $M$, and its normalized real-valued quantity is defined as:
\begin{equation}
V = (-1)^S \times (1.M) \times 2^{E - \text{Bias}} .
\end{equation}

In this expression, the ``1'' in $1.M$ is the hidden bit (Hidden Bit) that is automatically completed during hardware implementation. $\text{Bias}$ is the exponent bias value, which depends on the exponent bit-width $k$ and is calculated as: $\text{Bias} = 2^{k-1} - 1$. The workflow of a single-precision floating-point multiplier typically follows the five stages described below.

\textbf{Sign bit generation and operand preprocessing}: At the initiation stage of the operation, the hardware circuit first extracts the sign bits of the two operands $A$ and $B$. Since the positive and negative signs follow XOR logic, the resulting sign bit $S_{\text{res}}$ is determined by the following formula:
\begin{equation}
S_{\text{res}} = S_A \oplus S_B .
\end{equation}
Meanwhile, the logic unit checks the inputs for special values, such as zero, infinity, NaN, or subnormal numbers. When processing normalized numbers, the hardware extracts the mantissa field and prepends a ``1'' to its most significant bit, restoring it to the full significand for subsequent multiplication.

\textbf{Exponent accumulation and bias correction}: Since the exponents of the input operands both include the bias value, a direct summation of the two exponents would result in the bias being counted twice. Therefore, the calculation of the intermediate exponent $E_{\text{inter}}$ requires subtracting one bias value for correction:
\begin{equation}
E_{\text{inter}} = E_A + E_B - \text{Bias} .
\end{equation}
This stage also involves preliminary monitoring to detect whether the exponent might lead to potential overflow or underflow.

\textbf{Mantissa multiplication core}: The mantissa processing is the core part of the entire multiplier, accounting for the largest share of area and power consumption. Using a single-precision floating-point multiplier as an example, the hardware performs a multiplication of the two restored 24-bit unsigned fixed-point numbers to generate a 48-bit intermediate product ($IP$):
\begin{equation}
(1.M_A) \times (1.M_B) = IP[47:0] .
\end{equation}
Since the magnitude of each significand is within the range $[1.0, 2.0)$, the value of the product $IP$ must fall within the range $[1.0, 4.0)$.

\textbf{Normalization and exponent fine-tuning}: If the intermediate product $IP \ge 2.0$ (i.e., the 47th bit is 1), it indicates that a carry occurred, requiring right-normalization. Specific operations include shifting the product right by 1 bit and incrementing the intermediate exponent by 1, such that $E_{\text{final}} = E_{\text{inter}} + 1$. If the product is already within the range $[1.0, 2.0)$, no shift is required, and the exponent remains unchanged.

\textbf{Rounding treatment and overflow detection}: To fit the 48-bit intermediate result back into the 23-bit storage format defined by IEEE 754, the hardware must perform rounding according to specified modes, with the most common being ``round to nearest, ties to even''. At this stage, the system performs final exception determinations. Overflow: if $E_{\text{final}} > 254$, the result is set to $\pm\infty$; Underflow: if $E_{\text{final}} < 1$, the result is typically set to $\pm 0$.

After completing these five steps, the final generated sign bit, exponent, and rounded mantissa are reassembled into a floating-point result that complies with the IEEE 754 standard.

\subsection{Approximate Floating-Point Multiplier}

To overcome the prohibitive area and power overhead of conventional IEEE~754
floating-point multipliers while preserving the numerical dynamic range required
for training convergence, we propose a segmented approximate floating-point
multiplier integrated into the OpenACM framework. 

\begin{figure}[tb]
\centering
\includegraphics[width=0.9\linewidth]{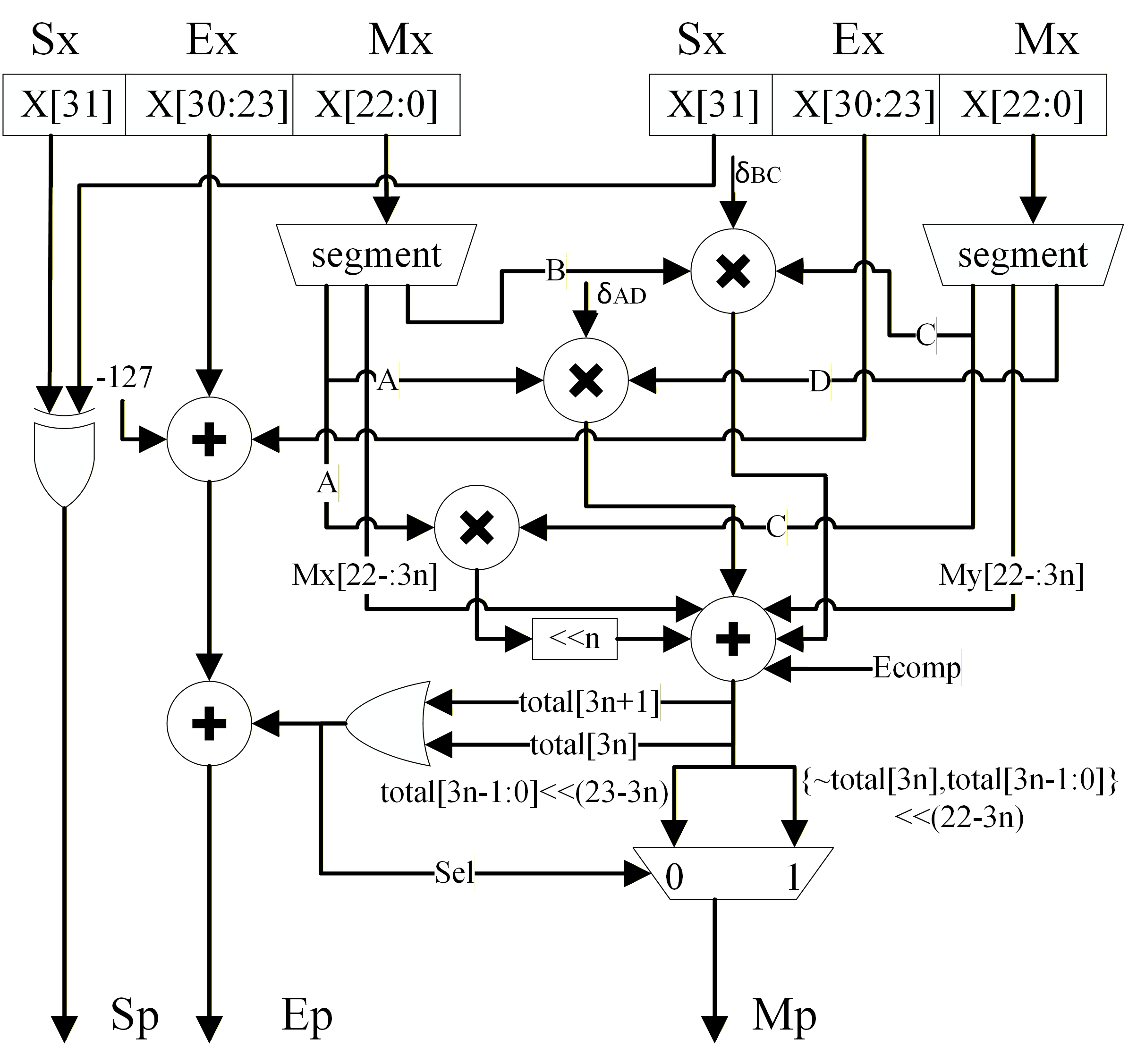}
\caption{Overall hardware architecture of the proposed approximate floating-point
multiplier.}
\label{fig:arch}
\vspace{-12pt}
\end{figure}

The overall hardware
architecture is illustrated in \autoref{fig:arch}, which highlights the key
modules including mantissa segmentation, conditional partial-product execution,
error compensation, shift-and-add accumulation, and normalization logic.

\begin{figure}[tb]
\centering
\includegraphics[width=0.95\linewidth]{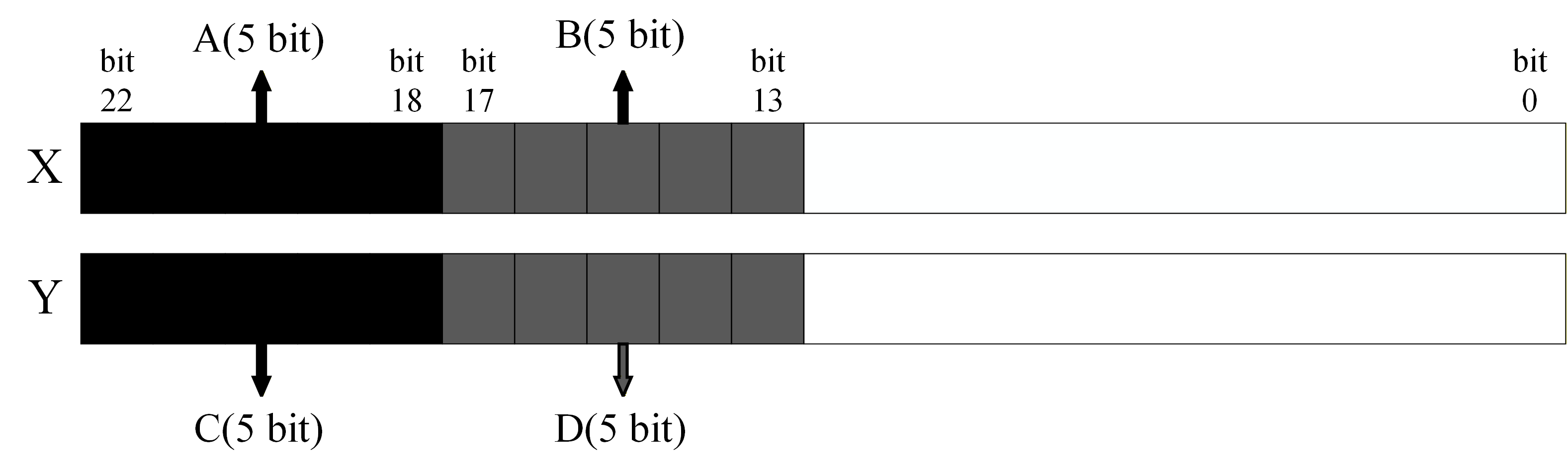}
\caption{Mantissa segmentation with configurable width $n$.}
\label{fig:segmentation}
\vspace{-12pt}
\end{figure}

For IEEE~754 single-precision floating-point numbers, the explicit mantissa
(excluding the hidden bit) has a width of 23 bits. In the proposed design, the
mantissas $M_X$ and $M_Y$ are partitioned into high-significance and
low-significance segments with a configurable segment width $n$, as shown in
\autoref{fig:segmentation}. Specifically, the mantissas can be expressed as
\begin{equation}
\begin{aligned}
M_X &= A \cdot 2^{-n} + B \cdot 2^{-2n}, \\
M_Y &= C \cdot 2^{-n} + D \cdot 2^{-2n},
\end{aligned}
\end{equation}
where $(A,C)$ represent the high-significance segments and $(B,D)$ denote the
low-significance segments. 

Expanding the exact mantissa product yields four
partial products: $AC$, $AD$, $BC$, and $BD$. Among these terms, $AC$ carries the highest weight and dominates the numerical
magnitude of the final result. Any perturbation in $AC$ would introduce a large
relative error; therefore, the $AC$ term is always computed exactly. The $AD$
and $BC$ terms have intermediate weights and are conditionally executed to
improve power efficiency. When the upper $(n-2)$ bits of $B$ or $D$ are all zero,
the corresponding value is sufficiently small such that executing a full
$n \times n$ multiplication provides limited accuracy improvement relative to
its energy cost. In these cases, the multiplier operation is bypassed.

Directly discarding the $AD$ or $BC$ term, however, would introduce noticeable
numerical bias. To mitigate this effect, a lightweight shift-based compensation
mechanism is employed. When conditional execution is disabled and both operands
are non-zero, the low-significance segment is approximated by a constant,
transforming the multiplication into a simple shift operation (e.g.,
$A \ll 1$). This strategy incurs negligible logic overhead while recovering more
than 50\% of the truncated error in practice. Special cases are also handled to
avoid pathological errors: if $A=0$ and $B,C \neq 0$, the $BC$ term is forced to
execute; similarly, if $C=0$ and $A,D \neq 0$, the $AD$ term is enforced.

The $BD$ term has the lowest weight. For example, when $n=5$, its weight is only
$1/1024$ of that of $AC$. Extensive empirical evaluation shows that omitting the
$BD$ term reduces average error metrics only marginally, while eliminating an
entire $n \times n$ multiplier array and its associated adder tree. This results
in an area reduction of approximately 6.8\% and a power reduction of 12.6\%,
making the omission of $BD$ highly cost-effective in area- and
power-constrained designs. Consequently, the approximate mantissa product can be
formulated as
\begin{equation}
M_X M_Y \approx (A \times C) \ll n
+ \delta_{AD}(A \times D)
+ \delta_{BC}(B \times C)
+ E_{\text{comp}},
\end{equation}
where $\delta_{AD}, \delta_{BC} \in \{0,1\}$ denote conditional execution flags,
and $E_{\text{comp}}$ represents the error compensation term.

\begin{figure}[tb]
\centering
\includegraphics[width=0.95\linewidth]{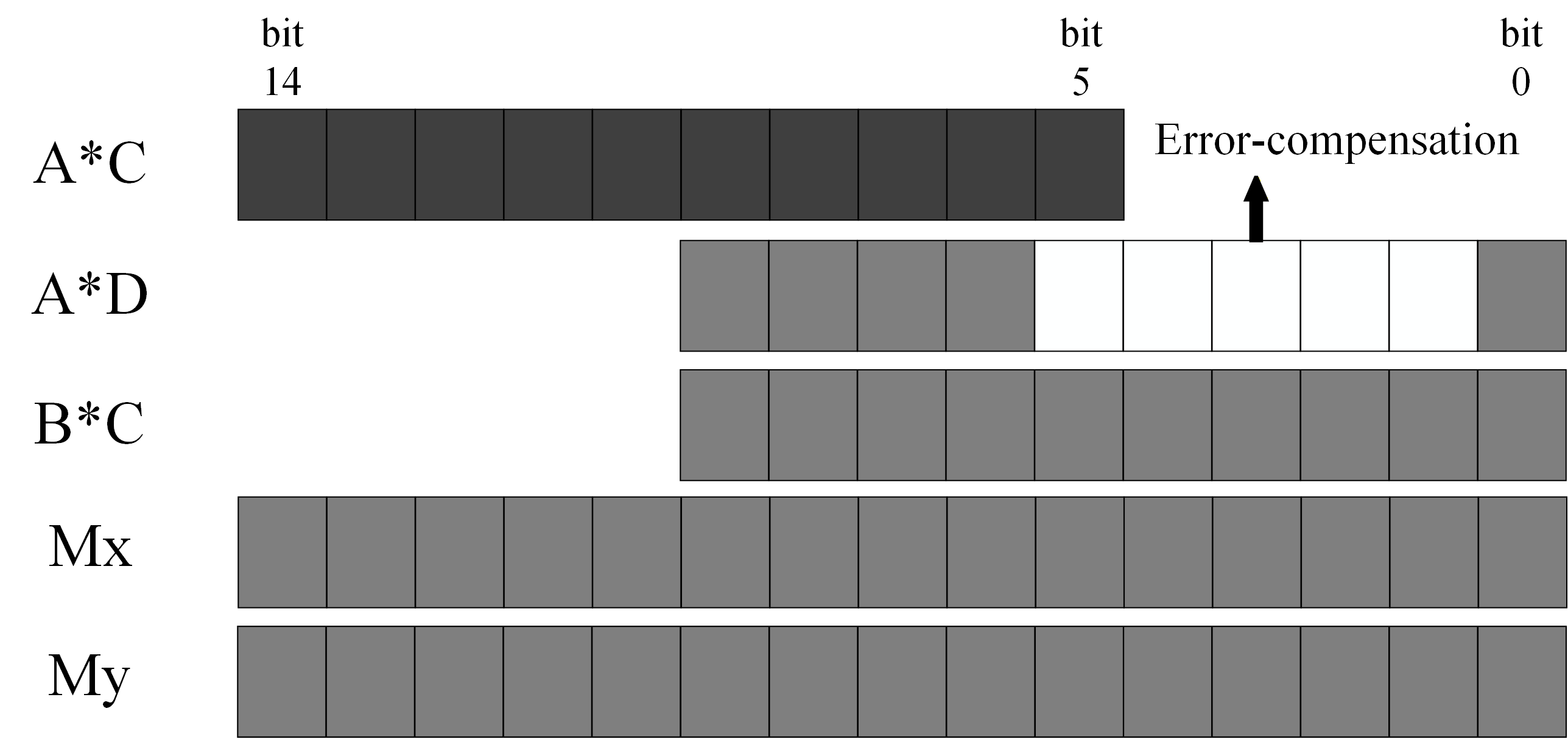}
\caption{Shift-and-add alignment and accumulation of partial products.}
\label{fig:shiftadd}
\vspace{-12pt}
\end{figure}

Before accumulation, partial products with different weights are aligned using
shift operations. To minimize hardware overhead while preserving high-order
accuracy, a significance-aware rounding strategy is adopted. Considering that
each partial product has a width of $2n$ bits and that the $AC$ term is left
shifted by $n$ bits, the accumulator width is optimized to $3n$ bits. The input
mantissas $M_X$ and $M_Y$ are correspondingly truncated to their upper $3n$ bits.
As illustrated in \autoref{fig:shiftadd}, the white regions indicate the bit
positions where compensation terms are inserted when conditional execution is
enabled.

\begin{figure*}[tb]
\centering
\includegraphics[width=0.9\linewidth]{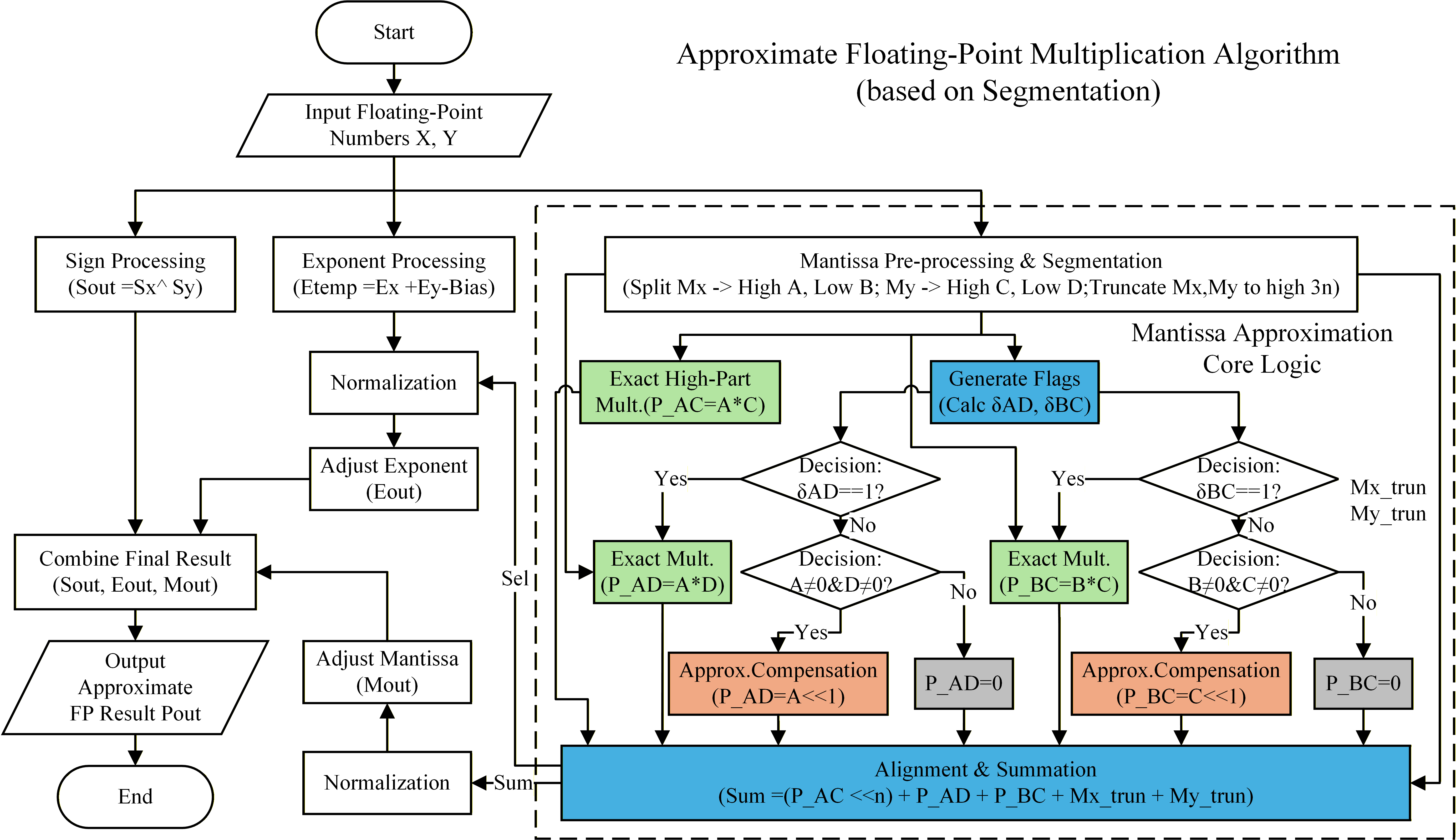}
\caption{Computation flow of the proposed mantissa-segmentation-based approximate floating-point multiplier.}
\label{fig:flow}
\vspace{-12pt}
\end{figure*}

Normalization is determined by the two most significant bits of the accumulated
result. If either of these bits is ``1'', the product exceeds the standard
mantissa range, triggering normalization by a left shift and an exponent
increment. During this process, bit-inversion logic is employed to correctly
handle the hidden bit defined by the IEEE~754 standard. If both bits are ``0'',
standard alignment is applied. The normalized mantissa is then zero-padded to
23 bits to conform to the FP32 format. The overall computation workflow of the
proposed approximate floating-point multiplier is summarized in
\autoref{fig:flow}.

By configuring the segment width $n$, the proposed architecture provides
fine-grained control over the accuracy--efficiency trade-off and supports
systematic design space exploration. Moreover, the framework is flexible enough
to accommodate floating-point formats with bit widths ranging from 16 to 32
bits, enabling seamless adaptation to diverse numerical requirements in
on-chip learning and fine-tuning tasks.
Furthermore, to better support applications with relaxed accuracy requirements but stringent area and power constraints, the proposed design provides an additional low-precision mode. This mode is motivated by the observation that the dominant high-order product contribution is mainly determined by the local correlation between the high-order bits of the inputs, and a bitwise AND operation can directly capture this joint-activation behavior. Therefore, it can serve as a first-order, low-complexity approximation of the high-order product without explicitly generating the full product term. Based on this approximation, the partial sum in this mode is composed only of the highest 
n-bit segments of the two input mantissas and their bitwise AND result, without involving any additional cross-product terms. As a result, the accumulation width is further reduced from 3n bits to n bits. This design significantly lowers the hardware complexity and switching activity of the mantissa multiplication core, thereby providing a more lightweight implementation option for area- and power-constrained scenarios.
\section{Experimental Results}
\label{sec:results}
\subsection{Post-Layout PPA Evaluation}

\begin{table}[tb]
\centering
\caption{Post-layout performance of integrating AFPMs to OpenACM, where SRAM access is the critical path (@100MHz, with 0.5pF output load.)}
\label{tab:sram_fp_performance}
\resizebox{\linewidth}{!}{%
\begin{tabular}{cccccccc}
\toprule
\multirow{2}{*}{SRAM} & \multirow{2}{*}{Multiplier Type} & \multirow{2}{*}{Delay (ns)} & \multicolumn{3}{c}{Area ($\mu$m\textsuperscript{2})} & \multirow{2}{*}{Power (W)} \\
\cmidrule{4-6}
 & & & Logic & SRAM & P\&R & \\
\midrule

% ================= 16x8 =================
\multirow{1}{*}{\begin{tabular}[c]{@{}c@{}}16 $\times$ 8\\ \end{tabular}}
& Exact & 5.22 & 705 & 7052 & 7757 & 1.73E--04 \\
\midrule

% ================= 32x16 =================
\multirow{2}{*}{\begin{tabular}[c]{@{}c@{}}32 $\times$ 16\\ \end{tabular}}
& Exact & 5.24 & 1829 & \multirow{2}{*}{16910} & 18739 & 5.29E--04 \\
& AC3-3 (Ours) & 5.24 & 1116 &  & 18026 & 3.08E--04\\
\midrule

% ================= 64x32 =================
\multirow{13}{*}{\begin{tabular}[c]{@{}c@{}}64 $\times$ 32\\ \end{tabular}}
& Exact     & 5.24 & 6268 & \multirow{13}{*}{48642} & 54910 & 2.32E--03 \\
& ACL5 (Ours)   & 5.24 & 1351 &  & 49993 & 4.16E--04 \\
& AC4-4 (Ours)   & 5.24 & 1945 &  & 50587 & 6.42E--04 \\
& AC5-5 (Ours)   & 5.24 & 2156 &  & 50798 & 7.72E--04 \\
& AC6-6 (Ours)   & 5.24 & 2568 &  & 51210 & 9.22E--04 \\
& MMBS5~\cite{afpm_mmbs} & 5.24 & 3134 &  & 51776 & 7.07E--04 \\
& MMBS6~\cite{afpm_mmbs} & 5.24 & 3171 &  & 51813 & 7.56E--04 \\
& MMBS7~\cite{afpm_mmbs} & 5.24 & 3329 &  & 51971 & 8.61E--04 \\
& CSS12~\cite{afpm_static_seg} & 5.24 & 2136 &  & 50778 & 6.42E--04 \\
& CSS14~\cite{afpm_static_seg} & 5.24 & 2312 &  & 50954 & 7.18E--04 \\
& CSS16~\cite{afpm_static_seg} & 5.24 & 2572 &  & 51214 & 8.01E--04 \\
& CSS18~\cite{afpm_static_seg} & 5.24 & 2846 &  & 51488 & 9.12E--04 \\
& NC~\cite{afpm_log_eafpu}    & 5.24 & 1360 &  & 50002 & 4.22E--04 \\
& LPC~\cite{afpm_log_eafpu}   & 5.24 & 1384 &  & 50026 & 4.33E--04 \\
& HPC~\cite{afpm_log_eafpu}   & 5.24 & 1658 &  & 50300 & 5.19E--04 \\
\bottomrule
\end{tabular}%
}\vspace{-12pt}
\end{table}

To evaluate hardware efficiency, post-layout performance and PPA analyses were conducted using the OpenROAD toolchain with the FreePDK45, under uniform constraints (100 MHz, 0.5 pF output load). The SRAM array organization was kept unchanged, and only the multiplier implementation was varied. \autoref{tab:sram_fp_performance} summarizes the post-layout results for different SRAM sizes and multiplier configurations. Compared with the exact multiplier, the proposed approximate designs achieve substantial reductions in logic area and power consumption without affecting the critical-path delay, as system timing is dominated by SRAM access. 
Specifically, ACL5 achieves the lowest hardware overhead among all evaluated designs, reducing logic area and power by 78.4\% and 82.1\%, respectively, relative to the exact multiplier, while also outperforming the aggressive NC baseline in hardware efficiency. In contrast, AC4-4 achieves substantial area and power savings while improving the mean relative error distance (MRED) and normalized mean error distance (NMED) by 80.4\% and 79.3\%, respectively, compared with HPC (see \autoref{tab:cnn}). These results indicate that the proposed architecture spans a wider accuracy-efficiency design space.

\subsection{Image Processing Evaluation}

\begin{table}[tb]
\centering
\caption{Comparison of different approximate multipliers for various image processing tasks.}
\label{tab:psnr_full}

% tighter column padding -> more room for font
\setlength{\tabcolsep}{3pt}
\renewcommand{\arraystretch}{1.05}

\begin{tabular*}{\linewidth}{@{\extracolsep{\fill}}%
>{\centering\arraybackslash}p{1.25cm}
>{\centering\arraybackslash}p{1.15cm}
c c c c c c
@{}}
\toprule
\multirow{2}{*}{Task} & \multirow{2}{*}{Test} & \multicolumn{6}{c}{Multiplier Type} \\
\cmidrule(lr){3-8}
& & AC4-4 & AC5-5 & AC6-6 & MMBS5 & CSS16 & HPC \\
\midrule
\multirow{3}{*}{\makecell{Image\\Blending}}
& Test1 & 61.59 & 74.74 & \textbf{86.44} & 56.16 & 73.40 & 45.97 \\
& Test2 & 59.35 & 72.51 & \textbf{84.50} & 55.07 & 71.03 & 43.96 \\
& Test3 & 62.59 & 75.48 & \textbf{86.43} & 56.04 & 74.58 & 46.48 \\
\midrule
\multirow{3}{*}{\makecell{Edge\\Detection}}
& Boat      & 82.97 & 96.31 & \textbf{109.80} & 73.08 & 93.85 & 74.25 \\
& Cameraman & 80.95 & 98.67 & \textbf{108.70} & 69.45 & 91.88 & 65.60 \\
& Jetplane  & 81.74 & 97.22 & \textbf{108.51} & 69.15 & 93.25 & 72.63 \\
\bottomrule
\multicolumn{8}{@{}l}{\footnotesize Test1: Lake~\&~Mandril; Test2: Jetplane~\&~Boat; Test3: Cameraman~\&~Lake.}
\end{tabular*}
\vspace{-12pt}
\end{table}

To evaluate practical performance, we tested the proposed approximate floating-point multipliers on representative image processing workloads, including image fusion and edge detection. These tasks are sensitive to accumulated arithmetic errors and thus provide an intuitive measure of numerical fidelity. Peak signal-to-noise ratio (PSNR) was adopted as the primary metric, and \autoref{tab:psnr_full} summarizes the results across different segmentation configurations and baseline designs. The proposed multipliers consistently achieve high PSNR across all tasks. As the segmentation width increases, PSNR improves nearly linearly, demonstrating fine-grained control over the accuracy–efficiency trade-off. In edge detection, PSNR exceeds 80 dB under multiple configurations, indicating visually indistinguishable results compared to exact floating-point computation. Compared with segmentation-based schemes such as MMBS and CSS, the proposed designs deliver substantial PSNR improvements, confirming that selective approximation in low-significance mantissa segments effectively suppresses perceptual error propagation. Moreover, the proposed architecture improves both accuracy and energy efficiency. AC4-4 reduces MRED by 52.7\% (to 1.38E–03) while saving 37.9\% logic area and 9.2\% power relative to MMBS5. Similarly, AC5-5 achieves a lower MRED (3.36E–04) with 16.1\% less area and 3.6\% lower power compared to CSS16. These results demonstrate that the proposed design enables precision-configurable and energy-efficient floating-point computation for DCiM arrays.

\subsection{CNN Inference Evaluation}
\begin{table}[tb]
\centering
\caption{Influence of approximate multipliers on ResNet-18.}
\label{tab:cnn}
\begin{tabular}{c c c c c}
\toprule
Multiplier Type & MRED & NMED & Top-1 & Top-5 \\
\midrule
Exact & -- & -- & 0.8715 & 0.9961 \\
ACL5 (Ours) & 4.16E--02 & 1.58E--04 & 0.8569 & 0.9943 \\
AC4-4 (Ours) & 1.38E--03 & 5.35E--06 & 0.8715 & 0.9960 \\
AC5-5 (Ours) & 3.36E--04 & 1.30E--06 & 0.8717 & 0.9960 \\
AC6-6 (Ours) & \textbf{8.29E--05} & \textbf{3.55E--07} & 0.8715 & 0.9960 \\
MMBS5~\cite{afpm_mmbs} & 2.92E--03 & 1.13E--05 & 0.8714 & 0.9961 \\
MMBS6~\cite{afpm_mmbs} & 1.14E--03 & 4.46E--06 & 0.8715 & 0.9961 \\
MMBS7~\cite{afpm_mmbs} & 5.04E--04 & 1.98E--06 & 0.8717 & 0.9961 \\
CSS12~\cite{afpm_static_seg} & 1.45E--03 & 5.67E--06 & 0.8718 & 0.9961 \\
CSS14~\cite{afpm_static_seg} & 7.08E--04 & 2.78E--06 & 0.8715 & 0.9961 \\
CSS16~\cite{afpm_static_seg} & 3.48E--04 & 1.37E--06 & 0.8717 & 0.9961 \\
CSS18~\cite{afpm_static_seg} & 1.73E--04 & 6.79E--07 & 0.8715 & 0.9961 \\
NC~\cite{afpm_log_eafpu}  & 4.37E--02 & 1.55E--04 & 0.8253 & 0.9918 \\
LPC~\cite{afpm_log_eafpu} & 2.81E--02 & 1.07E--04 & 0.8631 & 0.9947 \\
HPC~\cite{afpm_log_eafpu} & 7.06E--03 & 2.59E--05 & 0.8717 & 0.9961 \\
\bottomrule
\end{tabular}
\vspace{-12pt}
\end{table}

We further evaluate the impact of approximate floating-point multipliers on
deep neural network inference using ResNet-18 on the CIFAR-10 dataset. The
network is trained using standard floating-point arithmetic, and inference is
performed by replacing exact multipliers with approximate designs. This setup
isolates the effect of arithmetic approximation during inference and reflects
practical deployment scenarios. \autoref{tab:cnn} summarizes the inference accuracy and error metrics for
different configurations. 
For  AC4-4, AC5-5, and AC6-6, the proposed designs maintain very low numerical errors without observable degradation in Top-1 or Top-5 accuracy. For the more aggressive low-precision mode ACL5, the inference accuracy remains acceptable. Notably, compared with NC, ACL5 shows no obvious degradation in neural network accuracy while also achieving better logic area and power results. This indicates that the proposed architecture is suitable not only for high-accuracy approximate computing, but also for providing a more flexible and efficient option for applications with relaxed precision requirements through ACL5.

% =========================
\section{Conclusion}
\label{sec:conclusion}

This work introduces compiler-integrated floating-point multiplication for SRAM-based DCiM through the OpenACM framework. An exact IEEE~754-compliant multiplier is implemented as a baseline, and an accuracy-configurable approximate multiplier based on mantissa segmentation is proposed to enable fine-grained accuracy--PPA trade-offs. Experimental results show substantial area and power savings without timing penalty, while evaluations on image processing tasks and ResNet-18 inference confirm high numerical fidelity with negligible application-level accuracy degradation. By providing unified compiler-level support for both exact and approximate floating-point multipliers, this work establishes a practical and extensible foundation for future learning-oriented DCiM workloads.

\bibliographystyle{IEEEtran}
\bibliography{refs}
\end{document}